\documentclass[useAMS,usenatbib]{mnras}
\usepackage{amsmath}
\usepackage{graphicx}
\usepackage{grffile}
\usepackage{hyperref}
\usepackage{txfonts}
\usepackage{natbib}
\usepackage{bm}

\graphicspath{{images/}}

\def\apj{ApJ}
\def\apjl{ApJL}
\def\apjs{ApJ Suppl. Ser.}

\def\aap{A\&A}

\def\araa{ARAA}
\def\actaa{Acta Astron.}
\def\mnras{MNRAS}

\def\sovast{Sov. Astron.}

\newcommand{\di}{\mathrm{d}}
\newcommand{\therm}{a} 
\newcommand{\visc}{b} 
\newcommand{\centr}{\mathrm{c}}
\newcommand{\eff}{\mathrm{ef}}
\newcommand{\vs}{{\mathrm{v}_\mathrm{s}}}
\newcommand{\vphi}{{\mathrm{v}_\mathrm{\phi}}}
\newcommand{\R}{{\cal R}}
\newcommand{\hyp}[2]{{}_#1F_#2}
\newcommand{\tbound}{{\theta_*}}
\newcommand{\xbound}{{x_*}}
\newcommand{\Dx}{{\Delta x}}
\renewcommand{\Pr}{{\mathrm{Pr}}}
\newcommand{\ar}{{a_\mathrm{r}}}
\newcommand{\pvisc}{d}
\newcommand{\speedoflight}{c}
\newcommand{\taueff}{\hat{\tau}}
\newcommand{\magn}{{\mathrm{M}}}

\def\beq#1{\begin{equation}\label{#1}}
\def\eeq{\end{equation}}
\def\beqa#1{\begin{eqnarray}\label{#1}}
\def\eeqa{\end{eqnarray}}

\def\eqn#1{~(\ref{#1})}
\def\myfrac#1#2{\left(\frac{#1}{#2}\right)}

\def\comment#1{\relax}
\def\dfrac#1#2{\displaystyle\frac{#1}{#2}}

\title[Convection in accretion discs]{Convection in axially symmetric accretion discs with microscopic transport coefficients}
\author[Malanchev et al.] {K.L. Malanchev$^{1,2}$\thanks{E-mail: malanchev@sai.msu.ru}, K.A. Postnov$^{1,3}$, N.I. Shakura$^{1}$\\
$^{1}$ Sternberg Astronomical Institute, Moscow M.V. Lomonosov State University, Universitetskij pr., 13,  Moscow 119992, Russia\\
$^{2}$ Faculty of Physics, M.V. Lomonosov Moscow State University,
Leninskie Gory, Moscow 119991, Russia\\
$^{3}$ Institute of Theoretical and Experimental Physics, Moscow, Russia}	

\begin{document}

\date{Received ... Accepted ...}
\pagerange{\pageref{firstpage}--\pageref{lastpage}} \pubyear{2016}

\maketitle

\label{firstpage}
\begin{abstract}
The vertical structure of stationary thin  
accretion discs is calculated from the energy balance equation 
with heat generation 
due to microscopic ion viscosity $\eta$ and electron heat conductivity $\kappa$, both depending on temperature. In the optically thin discs 
it is found that 
for the heat conductivity increasing with temperature, the vertical temperature gradient exceeds the adiabatic value at some height, suggesting convective instability in the upper disc layer. 
There is a critical Prandtl number, $\Pr=4/9$, above which a Keplerian disc become fully convective. The vertical density distribution of optically thin laminar accretion discs as found from the hydrostatic equilibrium equation cannot be generally described by a polytrope but in the case of constant viscosity and heat conductivity. In the optically thick discs with radiation heat transfer, the vertical disc structure is found to be convectively stable for both absorption dominated and scattering dominated opacities, unless a very steep dependence of the viscosity coefficient on temperature is assumed. A polytropic-like structure in this case is found for Thomson scattering dominated opacity.
\end{abstract}


\section{Introduction}
\label{intro}

The origin of angular momentum transfer in accretion discs is the key issue in
accretion disc theory. The standard accretion disc theory \citep{1973SvA....16..756S,1973A&A....24..337S,1981ARA&A..19..137P}
assumes that turbulent viscosity, which can be parametrized by the dimensionless parameter $\alpha$, can be
responsible for the observed high mass accretion rate in compact X-ray sources, protoplanetary discs and in other astrophysical objects. 
From purely hydrodynamic point of view, 
Keplerian flows are stable against small perturbations according to the classical Rayleigh criterion, and 
various mechanisms giving rise to turbulence in Keplerian accretion discs have been discussed. 
For example, magnetorotational instability \citep{1991ApJ...376..214B,1998RvMP...70....1B} 
is thought to be responsible for turbulence in various astrophysical discs. 
Recently, in an attempt to search for purely hydrodynamic mechanisms of turbulence in shear flows, 
we have revisited the problem of turbulence appearance in thin Keplerian discs  
from small perturbations in non-ideal fluids  
with microscopic transport coefficients 
\citep[viscosity and heat conductivity;][]{2015MNRAS.448.3707S,2015MNRAS.451.3995S,2016arXiv160304878M}. By modal analysis, we have found that both in the simplest Boussinesq and
anelastic approximations of hydrodynamic equations, unstable axially symmetric modes can appear in the shear accretion flows, 
which may serve as seeds for turbulence  even in the absence of dynamically significant magnetic fields. In addition to traditional
modal analysis of small perturbations,  
non-modal analysis of transient perturbations can be a powerful tool for 
searching for possible mechanisms of hydrodynamic turbulence in accretion flows \citep[e.g.][]{2015PhyU...58.1031R}.  
 
In the modal analysis of perturbations in thin accretion discs, equations for small variations of dynamical variables (density, velocity, pressure) are formulated as a boundary value problem against a given background, which should be solution of unperturbed hydrodynamic
equations. In \citet{2015MNRAS.451.3995S} this problem was solved using a priori postulated polytropic 
vertical structure of the disc. Earlier it was conjectured \citep[e.g.][]{1998A&AT...15..193K} that the vertical structure
of stationary accretion $\alpha$-discs can admit an effective polytropic description. However, it is 
far from being obvious that this is the case if the microscopic transport coefficients (instead of the effective turbulent
viscosity prescription) are used in hydrodynamic equations. 

The purpose of this paper is to find solution of vertical structure of stationary shear accretion flows with 
microscopic transport coefficients --- dynamic ion viscosity $\eta$ and heat conductivity $\kappa$, which can 
be characterized by a dimensionless Prandtl number $\Pr$.
The ion viscosity in hot accretion disc was considered already by \citet{1978AcA....28..253P}
and was shown to be able to provide, in some cases, sufficiently high mass accretion rate through the disc. 
However, \citet{1978AcA....28..253P} did not calculate the vertical disc structure.  

In Section \ref{s:Sect2}, we find such a solution for optically thin laminar discs and show that 
with standard microscopic transport coefficients, for a given Prandtl number a convectively unstable layer
appear near the upper boundary of the disc, which can encompass the total disc height if the Prandtl number exceeds some critical value
($4/9$ for a Keplerian disc). The possibility of thermal convection in accretion discs was 
found earlier in shear-box calculations and discussed in the context of outward angular momentum transfer by \cite{2010MNRAS.404L..64L}.

In Section \ref{s:rad}, we consider the vertical disc structure with ion 
viscosity and radiative energy transfer, pertinent to optically thick accretion discs. 
Here, for completeness, we also calculate the 
vertical structure of the standard turbulized $\alpha$-discs. 
The knowledge of the vertical structure of such discs, in turn, is needed to calculate their radial structure, which is 
usually done by averaging accretion disc equations over the disc thickness.   
The optically thick discs with heat generation due to microscopic ion viscosity and radiation heat transfer are found
to be convectively stable for both absorption dominated (Kramer's opacity) and scattering dominated (Thomson opacity) cases.

\section{Optically thin discs with electron heat conductivity}
\label{s:Sect2}

We will consider axially symmetric non-magnetized accretion discs 
with microscopic transport coefficients --- dynamic viscosity $\eta$ and 
heat conductivity $\kappa$, which are functions of temperature $T$ only. In this section, the 
disc is assumed to be optically thin in the vertical direction. 
As in the standard accretion disc theory, the radial velocity $u_r$ is assumed to be
much smaller than the azimuthal velocity $u_\phi(r)$, which is a function of radius $r$ only.
We assume hydrostatic equilibrium ($u_z=0$) and geometrically thin discs, so the dynamical equations are reduced to
one equation for pressure $P$:
\beq{eq.hydrostat}
\frac{\partial P}{\partial z}=-\rho g_z=-\rho \Omega^2z\,.
\eeq 
Here $\Omega(r)$ is the angular velocity of the flow, which is determined by the 
gravitational potential.

The equation of state for a perfect gas is convenient to write in the form:
\beq{e:eos}
P=Ke^{s/c_V}\rho^\gamma\,,
\eeq 
where $K$ is a constant, $s$ is the specific entropy per particle, $c_V=c_P/\gamma=1/(\gamma-1)$ is the specific volume heat capacity, $c_P$ is the specific heat capacity at constant pressure
and $\gamma=c_P/c_V$ is the adiabatic index (5/3 for the perfect monoatomic gas). 
We will also use the equation of state in the form
\beq{e:eos1}
P=\frac{\rho{\cal R}T}{\mu}\,,
\eeq
where $\mu$ is the molecular weight and $\R$ is the universal gas constant.

The energy equation can be written
\beq{e:en}
\frac{\rho {\cal R} T}{\mu}\left[
\frac{\partial s}{\partial t}+(\bm{\mathrm{v}}\nabla)\cdot s
\right]=
\frac{\di E_\mathrm{visc}}{\di t\,\di V}-\nabla\cdot Q\,,
\eeq
where $\di E_\mathrm{visc} / \di t / \di V$ is the viscous dissipation rate per unit volume, 
${\cal R}$ is the universal gas constant, 
$\mu$ is the molecular weight, $T$ is the temperature, and terms on the right 
stand for the viscous energy production and the heat conductivity energy flux $Q$, 
respectively. The energy flux due to the heat conductivity is 
\beq{e:kappat}
\nabla\cdot Q=\nabla(-\kappa\nabla T).
\eeq
Note that both electrons and photons, and
at low temperatures neutral atoms, can contribute to the heat conductivity.
The viscous heat generation in the unperturbed axially symmetric shear flow with angular velocity $\Omega(r)$ is 
\beq{e:qvisc}
\frac{\di E_\mathrm{visc}}{\di t\,\di V}=\eta\left[r\frac{d\Omega}{dr}\right]^2\,.
\eeq 

We are searching for vertical structure of the disc, so only $T(z)$ dependence is 
important. Noticing that in the axially symmetric case with small radial velocity 
$(\bm{\mathrm{v}} \nabla)\cdot s=0$, the energy equation can be cast to the form:
\begin{equation}
    P \frac{\partial s}{\partial t} =
        \frac{\partial}{\partial z} \left( \kappa(T) \, \frac{\partial T}{\partial z} \right) +
        \eta(T) \, r^2 \, \left( \frac{\di\Omega}{\di r} \right)^2 \,.
\label{eq.energy}
\end{equation}

\subsection{Temperature distribution from the energy equation}
\label{s:T(z)}

In the steady-state case, the energy equation (\ref{eq.energy}) enables us to calculate
the vertical temperature distribution $T(z)$ in a flow with given $\Omega(r)$ and 
microscopic transport coefficients $\kappa(T)$ and $\eta(T)$. 

It is convenient to introduce the dimensionless temperature $\theta$ and vertical coordinate $x$:
\begin{eqnarray}
    \theta &\equiv& \frac{T}{T_\centr} \,, \label{eq.t} \\
    x &\equiv& \frac{z}{z_0} \,, \label{eq.x}
\end{eqnarray}
where $T_\centr$ is the temperature in the disc symmetry plane and $z_0$ is its semithickness.

The central temperature $T_\centr$ can be expressed in terms of the adiabatic sound velocity $\vs$ in the disc symmetry plane:
\begin{equation}
    \frac{\R T_\centr}{\mu} = \frac{\vs^2}{\gamma} = \frac1{\gamma} \left(\frac{\vs}{\vphi}\right)^2 \Omega^2 r^2 \,,
\label{eq.T_centr}
\end{equation}
where $\vs/\vphi$ is the parameter of the model. 
It is also convenient to introduce the dimensionless parameter $\xi$ characterizing the relative thickness of the flow:
\begin{equation}
    \xi \equiv \left(\frac{\vs}{\vphi}\right)^{-1} \left(\frac{z_0}{r}\right).
\label{eq.xi}
\end{equation}

The dynamical viscosity $\eta$ and heat conductivity $\kappa$ are assumed to be power-law functions of temperature:
\begin{eqnarray}
    \kappa(\theta) &=& \kappa_\centr \theta^\therm \,, \label{eq.kappa} \\
    \eta(\theta) &=& \eta_\centr \theta^\visc \,, \label{eq.eta}
\end{eqnarray}
where values in the disc symmetry plane are related by the Prandtl number:
\begin{equation}
    \Pr \equiv \frac{\eta_\centr}{\kappa_\centr} \frac{ \R \, c_P }{\mu} \,.
\label{eq.Prandtl}
\end{equation}
Below we shall consider
only the physically relevant case of non-negative $a\ge 0$, $b\ge 0$.  

With these notations, equation~\eqref{eq.energy} can be written in the dimensionless form:
\begin{equation}
    P \frac{\partial s}{\partial t} =
        \frac{\kappa(\theta) \, T_\centr}{z_0^2}
        \left[
            \frac{\partial^2 \theta}{\partial x^2} +
            \frac{\therm}{\theta} \left( \frac{\partial t}{\partial x} \right)^2 +
            \theta^{\visc-\therm} \, \frac{\Pr \, \gamma \, \xi^2}{c_P} \left(\frac{\di\log\Omega}{\di\log r}\right)^2
        \right] \,.
\label{eq.energy_dimless}
\end{equation}

Let us introduce the dimensionless coefficient $B$:
\begin{equation}
    B \equiv \frac{\Pr \, \xi^2}{c_V} \left(\frac{\di\log\Omega}{\di\log r}\right)^2 \,.
\label{eq.B}
\end{equation}

For stationary flows $\partial / \partial t = 0$, and then  
the stationary dimensionless temperature vertical distribution $\theta(x)$ is a solution of the non-linear differential equation:
\begin{equation}
    \frac{\partial^2 \theta}{\partial x^2} +
        \frac{\therm}{\theta} \left( \frac{\partial \theta}{\partial x} \right)^2 + 
        B \theta^{\visc-\therm} =
        0 \,.
\label{eq.difeq}
\end{equation}

By vertical symmetry of the disc relative to the disc plane ($x = z = 0$) we have 
\begin{equation}
    \left. \frac{\partial \theta}{\partial x} \right|_{x=0} = 0 \,.
\label{eq.bound0_dir}
\end{equation}
The second boundary condition comes from the definition of the dimensionless temperature:
\begin{equation}
    \theta|_{x=0} = 1\ \,.
\label{eq.bound0_val}
\end{equation}

The differential equation~\eqref{eq.difeq} can be simplified by introducing new variable $u(\theta) \equiv (\partial \theta / \partial x)^2$:
\begin{equation}
 \begin{split}
    &\dfrac{\di u(\theta)}{\di \theta} + 2\therm \dfrac{u(\theta)}{\theta} + 2B\,\theta^{\visc-\therm} = 0 \,, \\
    &u|_{\theta=1} = 0 \,.
 \end{split}
\label{eq.difeq_dudt}
\end{equation}
This is an ordinary differential equation and its solution is
\begin{equation}
    u(\theta) = \left( \frac{\partial \theta}{\partial x} \right)^2 =
        \frac{2B}{\therm+\visc+1} \left( \frac{1 - \theta^{\therm+\visc+1}}{\theta^{2\therm}} \right) \,.
\label{eq.u}
\end{equation}
Taking square root of the last expression yields 
another ordinary differential equation that becomes linear if $x$ is a function of $\theta$:
\begin{eqnarray}
    \frac{\partial x(\theta)}{\partial \theta} &=&
        - \sqrt{ \frac{\therm+\visc+1}{2B} } \frac{\theta^\therm}{\sqrt{1-\theta^{\therm+\visc+1}}} \,, \label{eq.dx_dt} \label{eq.dx_dt} \\
    x|_{\theta=1} &=& 0 \,, \label{eq.boundary_x}
\end{eqnarray}
where the minus sign in the right-hand side of the first relation shows that temperature decreases with height.

The solution of the differential equation~\eqref{eq.dx_dt} with boundary condition~\eqref{eq.boundary_x} is
\begin{equation}
 \begin{split}
    x(\theta) & = 
        \sqrt{\frac{\therm+\visc+1}{2B}} \frac1{\therm+1} \times \\
        & \times \left[
            \hyp{2}{1}\left(\frac12, \frac{\therm+1}{\therm+\visc+1}; \frac{2\therm+\visc+2}{\therm+\visc+1}; 1\right) - \right.\\
            &\left. - \, \theta^{\therm+1} \hyp{2}{1}\left(\frac12, \frac{\therm+1}{\therm+\visc+1}; \frac{2\therm+\visc+2}{\therm+\visc+1}; \theta^{\therm+\visc+1}\right)
        \right] \,,
 \end{split}
\label{eq.x_t}
\end{equation}
where $\hyp{2}{1}$ is the Gaussian hypergeometric function.

\subsection{Convection instability of the background solution}

Equation~\eqref{eq.dx_dt} implies that for $\therm > 0$ (heat conductivity increasing with temperature), the 
vertical temperature gradient $\partial \theta / \partial x$ goes to negative infinity when $\theta$ goes to zero.
This suggests that if the surface temperature of the flow is small enough, its upper layer of the flow is convective.
To see this, apply the local Schwarzschild criterion for convection:
\begin{equation}
    \left|\frac{\partial \theta}{\partial x}\right| \geq  \left|\left( \frac{\partial \theta}{\partial x} \right)_\mathrm{ad}\right|,
\label{eq.schwarzschild}
\end{equation}
where the right-hand side is the adiabatic temperature gradient:
\begin{equation}
    \left( \frac{\partial \theta}{\partial x} \right)_\mathrm{ad} =
        \frac{z_0}{T_\centr} \left( \frac{\partial T}{\partial z} \right)_\mathrm{ad} =
        \frac{z_0}{T_\centr} \frac{\partial P}{\partial z} \frac{T}{P} \left( \frac{\di \log T}{\di \log P} \right)_\mathrm{ad} \,,
\label{eq.dtdx_ad}
\end{equation}
and $(\di \log T / \di \log P)_\mathrm{ad}=1/c_P$ is the adiabatic logarithmic temperature gradient for perfect gas.


Plugging equations~\eqref{eq.T_centr}, \eqref{eq.hydrostat} and~\eqref{e:eos1} 
into~\eqref{eq.dtdx_ad} yields:
\begin{equation}
    \left( \frac{\partial \theta}{\partial x} \right)_\mathrm{ad} = - \frac{\xi^2}{c_V} x \,.
\end{equation}
With the above relation, the criterion~\eqref{eq.schwarzschild} for the temperature gradient 
can be used to find the boundary $\xbound$ between laminar and convective layers:
\begin{equation}
    -\left.\frac{\partial \theta}{\partial x}\right|_{x=\xbound} = \frac{\xi^2}{c_V} \xbound \,.
\label{eq.t_bound_crit}
\end{equation}
The boundary between the layers $\xbound$ and corresponding temperature $\tbound$ can be found numerically.

\subsection{The critical Prandtl number}
\label{s:Prcr}

For some parameters, the Schwarzschild criterion~\eqref{eq.schwarzschild} 
for the background solution~\eqref{eq.x_t} is satisfied at any $1\ge x\ge 0$. To see this, 
consider the disc symmetry plane and expand the left-hand side of equation~\eqref{eq.t_bound_crit} in Taylor series
about the point $x=0$:
\begin{equation}
    -\left.\frac{\partial \theta}{\partial x}\right|_{x = \Dx} =
        - \left.\frac{\partial \theta}{\partial x}\right|_{x = 0}
        - \left.\frac{\partial^2 \theta}{\partial x^2}\right|_{x = 0} \Dx
        + O(\Dx^2) \,.
\end{equation}
Here $\Dx \ll 1$, the first term in the right-hand side vanishes by the boundary condition 
\eqref{eq.bound0_dir} and the second term is $B \, \Dx$ by~\eqref{eq.difeq} at $\theta=1$.

Therefore, about the disc symmetry plane 
the convection condition~\eqref{eq.schwarzschild} can be written as 
\begin{equation}
    B \, \Dx \geq \frac{\xi^2}{c_V} \Dx \,.
\end{equation}
Using the definition of $B$~\eqref{eq.B} 
we arrive at the condition on the Prandtl number for convection to occur 
across the entire disc height:
\begin{equation}
    \Pr \geq \Pr_\mathrm{crit}=\left(\frac{\di\log\Omega}{\di\log r}\right)^{-2} \,.
\label{eq.critical_Prandtl}
\end{equation}
For a Keplerian flow with $\Omega \sim r^{-3/2}$ we find $\Pr_\mathrm{crit} = 4/9$.

For fully ionized gas without magnetic field $a = b = 5/2$ and the Prandtl number $\Pr = 0.052$ \citep{1962pfig.book.....S} 
and the convection layer is narrow (see Fig.~\ref{fig.theta_x}).
However, if there is a small (dynamically unimportant) magnetic field in a plasma, 
the motion of electrons becomes bounded thus decreasing the heat conductivity $\kappa_\magn$:
\begin{equation}
    \kappa_\magn = \frac{\kappa}{1 + (\omega_\mathrm{c} \tau)^2} \,,
\label{eq.kappa_magn}
\end{equation}
where $\omega_\mathrm{c}$ is the cyclotron frequency and $\tau$ is the characteristic collision time.
Therefore, in such a slightly magnetized plasma 
the Prandtl number~\eqref{eq.Prandtl} can increase and the convection layer widens.

In the case of neutral gas, $a = b = 1/2$ and $\Pr = 2/3$ in the simplest molecular model  \citep{hirschfelder1954molecular}, 
and the Keplerian accretion disc turns out to be fully convective.

\subsection{Structure of the convective layer}

We will not solve the energy equation in the convective layer 
and simply will assume that the 
temperature gradient in this layer matches 
the adiabatic gradient~\eqref{eq.dtdx_ad} (see Section \ref{s:discus} for the discussion):
\begin{equation}
    \frac{\partial \theta}{\partial x} = - \frac{\xi^2}{c_V} x(\theta) \quad \text{for} \quad x \geq \xbound \,.
\label{eq.conv_dt_dx}
\end{equation}
Integrating this differential equation with the boundary condition $x(\tbound) = \xbound$ yields:
\begin{eqnarray}
    &x(\theta) = \sqrt{ \dfrac{2 c_V}{\xi^2} (\tbound - \theta) + \xbound^2 }
        \quad & \text{for} \quad \theta \leq \tbound \,, \label{eq.conv_x_t} \\
    &\theta(x) = \tbound - \dfrac{\xi^2}{2 c_V} (x^2 - \xbound^2)
        \quad & \text{for} \quad x \geq \xbound \,. \label{eq.conv_t_x}
\end{eqnarray}

\begin{figure}
 \centering
 \includegraphics[width=0.5\textwidth]{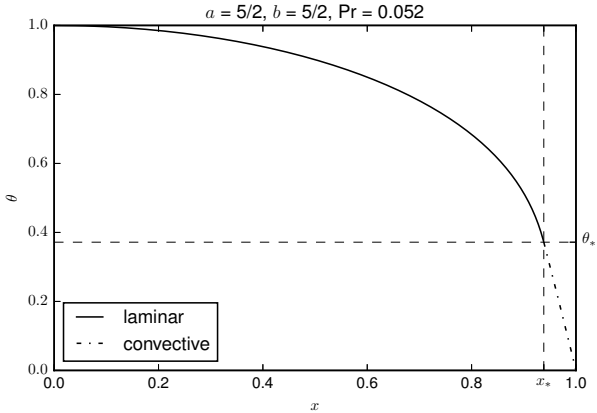}
 \caption{Vertical distribution of the dimensionless temperature $\theta$ in a thin Keplerian disc with 
ion viscosity and heat conductivity ($a = b = 5/2$, $\Pr = 0.052$). The solid line shows 
the temperature distribution in the laminar bottom layer of the disc ($x < \xbound$ and $\theta > \tbound$) 
with $\theta$ from equation~\eqref{eq.x_t}. The dash-dotted line shows 
the temperature distribution in the convective upper layer with $\theta$ 
from  equation~\eqref{eq.conv_t_x}. Here the boundary values are $\xbound \approx 0.938$ and $\tbound \approx 0.372$, the dimensionless parameter $\xi \approx 3.055$.}
\label{fig.theta_x}
\end{figure}

Fig.~\ref{fig.theta_x} presents the temperature distribution $\theta(x)$ for a 
fully ionized thin Keplerian disc with microscopic ion viscosity and electron heat conductivity 
characterized by the coefficients $a = b = 5/2$ and the Prandtl number $\Pr = 0.052$ \citep{1962pfig.book.....S}.
The bottom laminar layer is shown by the solid line. At the height $\xbound$, the temperature gradient $d\theta/dx$ becomes
superadiabatic, and the layer above this height the disc is prone to convection instability (the dash-dotted line).  

\subsection{Thickness of the accretion flow}

To find the upper boundary of the disc, we assume that the temperature vanishes at the disc surface:
\begin{equation}
    x|_{\theta=0} = z|_{\theta=0}/z_0 = 1 \,.
\label{eq.bound1_val}
\end{equation}
This condition enables us to find the dimensionless disc thickness $\xi$.
We consider two models of the disc structure: laminar flow and flow with laminar and convective layers.

\subsubsection{Laminar flow}
\label{s:laminar_flow}
In this section we will assume that even if the temperature gradient in the layer exceeds the 
adiabatic one, the temperature distribution follows the background solution~\eqref{eq.x_t}.
Then from the boundary condition~\eqref{eq.bound1_val} and expression~\eqref{eq.x_t} we find
\begin{equation}
    \xi = \sqrt{\frac{\therm+\visc+1}{2 \frac{\Pr}{c_V} \left(\frac{\di\log\Omega}{\di\log r}\right)^2}}
        \frac1{\therm+1} \,
        \hyp{2}{1}\left(\frac12, \frac{\therm+1}{\therm+\visc+1}; \frac{2\therm+\visc+2}{\therm+\visc+1}; 1\right) \,.
\label{eq.xi_laminar}
\end{equation}

The solution~\eqref{eq.x_t} itself takes the form
\begin{equation}
    x(\theta) = 1 - \theta^{\therm+1} \frac{ \hyp{2}{1}\left(\frac12, \frac{\therm+1}{\therm+\visc+1}; \frac{2\therm+\visc+2}{\therm+\visc+1}; \theta^{\therm+\visc+1}\right) }{ \hyp{2}{1}\left(\frac12, \frac{\therm+1}{\therm+\visc+1}; \frac{2\therm+\visc+2}{\therm+\visc+1}; 1\right) } \,.
\label{eq.x_t_simpl}
\end{equation}

Note that in the particular case where $\therm = \visc = 0$ and $\Pr~<~\Pr_\mathrm{crit}$
the criterion for convection to arise~\eqref{eq.schwarzschild} does not meet for any $1~\geq~x~\geq~0$.
In this case equations~\eqref{eq.xi_laminar} and~\eqref{eq.x_t_simpl} transform to
\begin{eqnarray}
    \xi &=& \sqrt{\frac2{\frac{\Pr}{c_V} \left(\frac{\di\log\Omega}{\di\log r}\right)^2}} \,, \label{eq.poly_xi} \\
    x(\theta) &=& \sqrt{1-\theta} \,, \label{eq.poly_x_t} \\
    \theta(x) &=& 1 - x^2 \,. \label{eq.poly_t_x}
\end{eqnarray}

\subsubsection{Two-layer flow}
In the case of two-component accretion flow (i.e. at Prandtl numbers below the critical one, see
Section \ref{s:Prcr}), the convection layer lies above the laminar one and the 
temperature vanishes at the convection zone surface.
Then expression~\eqref{eq.conv_x_t} and the boundary condition~\eqref{eq.bound1_val} imply
\begin{equation}
    \xi = \sqrt{ 2c_V \frac{\tbound}{1-\xbound^2} } \,,
\label{eq.conv_xi}
\end{equation}
where $\xbound$ and $\tbound$ can be found numerically from~\eqref{eq.t_bound_crit}.

\subsection{Density distribution from the hydrostatic equilibrium equation}

The solution $T(z)$ obtained above allows us to calculate the vertical density 
distribution $\rho(z)$ from the hydrostatic equilibrium equation~\eqref{eq.hydrostat}. 

Let us introduce the dimensionless density $\lambda$:
\begin{equation}
    \lambda \equiv \frac{\rho}{\rho_\centr} \,.
\label{eq.A.dimless_density}
\end{equation}
Using the equation of state~\eqref{e:eos1} 
and equations~\eqref{eq.t}, \eqref{eq.x}, \eqref{eq.T_centr}, and~\eqref{eq.xi}, 
the hydrostatic equilibrium equation~\eqref{eq.hydrostat} can be written as:
\begin{equation}
    \frac{\partial\lambda}{\partial x} \theta =
        - \lambda \frac{\partial\theta}{\partial x}
        - \lambda \, \xi^2 \, \gamma \, x \,.
\end{equation}
Dividing this equation through by $(\partial \theta / \partial x) \theta \lambda$, 
we obtain a linear differential equation for the function $\ln\lambda(\theta)$:
\begin{equation}
    \frac{\partial\ln\lambda}{\partial\theta} =
        - \frac1{\theta}
        - \xi^2 \, \gamma \, \frac{x(\theta)}{\theta} \frac{\partial x(\theta)}{\partial \theta} \,.
\label{eq.drho_dt}
\end{equation}

For $\therm > 0$ the solution of this equation can be straightforwardly found analytically for the laminar layer~\eqref{eq.x_t} through generalized 
hypergeometric functions $\hyp{3}{2}$ and gamma-function.
For the convective layer using equations~\eqref{eq.conv_dt_dx},~\eqref{eq.conv_x_t} and~\eqref{eq.conv_xi} solution of~\eqref{eq.drho_dt} can be obtained:
\begin{equation}
    \lambda \sim \theta^{c_V} \quad \text{for} \, \theta \leq \tbound \,,
\label{eq.conv_rho_t}
\end{equation}
which is manifestly polytropic.
%

The vertical density distribution in the laminar layer generally is not a polytrope but in one special case with constant transport coefficients
$\therm = \visc = 0$ and $\Pr < \Pr_\mathrm{crit}$ when no convective layer appears and the optically thin disc is fully laminar (see Section~\ref{s:laminar_flow} above). In this case equation~\eqref{eq.drho_dt} takes the form:
\begin{equation}
    \frac{\partial\ln\lambda}{\partial\theta} = \frac1{\theta} \left( \frac{\xi^2 \, \gamma}{2} - 1 \right) \,.
\label{eq.A.poly_drho_dt}
\end{equation}
The solution of this equation is
\begin{equation}
    \lambda = \theta^{\xi^2 \, \gamma / 2 - 1} = (1-x^2)^{\xi^2 \, \gamma / 2 - 1} \,.
\label{eq.A.rho_t}
\end{equation}
Thus, a laminar disc with constant heat conductivity and dynamical viscosity 
$\therm = \visc = 0$ has a polytropic vertical structure with the polytrope index $n$:
\beq{eq.A.xi_n}
    n =\dfrac{\xi^2 \, \gamma}{2} - 1 \,. \
\eeq
Plugging~\eqref{eq.poly_xi} into~\eqref{eq.A.xi_n} with account for~\eqref{eq.critical_Prandtl} leads to the convection 
stability condition $n > c_V$, which is, of course, 
well known for polytropes and can easily be derived from \eqn{e:eos}.

\section{Radiative heat conductivity}
\label{s:rad}
In this section we turn to the case of the radiative heat conductivity in an optically thick shear flow, which is relevant to standard accretion discs. 

\subsection{Vertical structure equations}
The radiative energy transfer equation in the diffusion approximation reads:
\begin{equation}
\label{e:radendiff}
    \frac{\partial \, \ar \speedoflight T^4}{\partial z} = - 3 \, \varkappa(\rho,T) \, \rho \, Q \,,
\end{equation}
where $\ar$ is the radiation constant, $\speedoflight$ is the  is the speed of light, $Q$ is the radiation flux, and $\varkappa(\rho,T)$ is the Rosseland opacity.
We will assume the opacity coefficient in the power-law form of density and temperature:
\begin{equation}
    \varkappa(\rho, T) = \varkappa_0  \frac{\rho^\varsigma}{T^\psi} = \varkappa_\centr \frac{p^\varsigma}{\theta^{\psi+\varsigma}} \,,
\label{eq.varkappa}
\end{equation}
where $p \equiv P / P_\centr$ is the dimensionless pressure.
For example, for Kramer's free-free opacity $\varsigma = 1$ and $\psi = 7/2$, 
for free-free opacity with solar abundance in the temperature range $10^4-10^6$~K and densities $10^{-10}-10^{-6}$~g~cm$^{-3}$
$\varsigma \approx 1$ and $\psi \approx 2.5$ \citep{1994ApJ...427..987B}, and 
for Thomson scattering $\varsigma = \psi = 0$.

To be able to compare our equations and results with the standard $\alpha$-disc model,
in this section we will parametrize the viscosity coefficient in the form different from~\eqref{eq.eta} :
\begin{equation}
    \eta = \eta_\centr \theta^\visc p^\pvisc \,.
\label{eq.eta_tp}
\end{equation}
For dynamic viscosity considered up to now $\pvisc = 0$, and for turbulized $\alpha$-disc 
$\eta_\centr = \alpha P_\centr / (-r \di\Omega / \di r)$, $\visc = 0$ and $\pvisc = 1$.

Introduce the dimensionless mass coordinate $\sigma$:
\begin{equation}
\label{e:sigmadef}
    \sigma(z) \equiv \frac{ \int_0^z{\rho \, \di z} }{\Sigma} \,,
\end{equation}
where $\Sigma \equiv \int_0^{z_0}{\rho \, \di z}$ is half the surface density of the flow.

The full system of differential equations for the disc vertical structure can be written
as a function of the mass coordinate $\sigma$ in the form similar to that used in \citet{1998A&AT...15..193K}:
\begin{equation}
 \begin{array}{llll}
    \dfrac{\partial p}{\partial\sigma} &=& - \Pi_1 \, \Pi_2 \, x \,, \quad
        &\Pi_1 \equiv \dfrac{ \Omega^2 \, z_0^2 \, \mu}{\R \, T_\centr} = \dfrac{\xi^2}{\gamma} \,; \\[4mm]
    \dfrac{\partial x}{\partial\sigma} &=& \Pi_2 \, \dfrac{\theta}{p} \,, \quad
        &\Pi_2 \equiv \dfrac{\Sigma}{z_0\, \rho_\centr} \,; \\[4mm]
    \dfrac{\partial q}{\partial \sigma} &=& \Pi_3 \, \theta^{\visc+1} p^{\pvisc-1} \,, \quad
        &\Pi_3 \equiv \eta_\centr \dfrac{\Sigma}{Q_0 \, \rho_\centr} \left(r \dfrac{\di\Omega}{\di r}\right)^2 \,; \\[4mm]
    \dfrac{\partial \theta}{\partial\sigma} &=& - \Pi_4 \, q \dfrac{p^{\varsigma}}{\theta^{\psi+\varsigma+3}} \,,  \quad
        &\Pi_4 \equiv \dfrac{3}{16} \, \varkappa_\centr \, \Sigma \left(\dfrac{ T_\eff}{T_\centr}\right)^4 \,;
 \label{eq.Pis}
 \end{array}
\end{equation}
where $Q_0 \equiv \ar \speedoflight \, T_\eff^4 / 4$ is the radiative flux at the surface of the flow, 
$T_\eff$ is the effective temperature  
(assumed to be equal to the surface temperature), $q \equiv Q / Q_0$ is dimensionless 
energy flux, and $\Pi_{1..4}$ are dimensionless constants. 
The first equation in~\eqref{eq.Pis} is the hydrostatic equation~\eqref{eq.hydrostat}, 
the second equation in~\eqref{eq.Pis} follows from the definition of the mass coordinate~\eqref{e:sigmadef}, 
the third equation in~\eqref{eq.Pis} is the viscous energy generation equation~\eqref{e:qvisc}, and 
the fourth equation in~\eqref{eq.Pis} is the radiation energy diffusion equation~\eqref{e:radendiff}. 
In total, four unknown functions ($p, x, q, \theta$) and four dimensionless constants $\Pi_{1..4}$ 
are to be determined from the system~\eqref{eq.Pis} subjected to eight boundary conditions.

\subsection{Boundary conditions}
\label{s:boundcond}

Six boundary conditions immediately follow from the definitions of the dimensionless unknown functions.
Four boundary conditions in the disc symmetry plane ($\sigma = 0$) reads:
\begin{equation}
    p_{\sigma=0} = 1\,, \quad
    x_{\sigma=0} = 0\,, \quad
    q_{\sigma=0} = 0\,, \quad
    \theta_{\sigma=0} = 1 \,.
\label{eq.bounds_sigma0}
\end{equation}
Another two boundary conditions are found at the surface of the accretion flow:
\begin{equation}
    x_{\sigma=1} = 1\,, \quad
    q_{\sigma=1} = 1\,.
\label{eq.bounds_sigma1}
\end{equation}

The remaining two boundary conditions for 
surface values of the dimensionless pressure $p$ and  temperature $\theta$ are 
determined by the location of photosphere which depends on the opacity law.
In this work we examine absorption dominated and scattering dominated cases.

\subsubsection{Absorption-dominated atmosphere}
\label{s:absorption}
In the upper layer of the flow the energy release is small, therefore to locate the photosphere we can use the Eddington approximation:
\begin{equation}
    \frac{T}{T_\eff} = \left( \frac{1+\frac32 \tau}{2} \right)^{1/4},
\end{equation}
where $\tau$ is the optical depth counted from the observer to the photosphere.

Setting the photosphere boundary $\sigma = 1$ at the point where $\tau = 2/3$ and $T = T_\eff$,
the boundary condition for the dimensionless temperature $\theta$ is
\begin{equation}
    \theta|_{\sigma = 1} = \left( \frac{16}{3} \frac{\Pi_4}{\tau_0} \right)^{1/4} \,,
\label{eq.ff_bound_t1}
\end{equation}
where $\tau_0 \equiv \varkappa_\centr \, \Sigma$ is the dimensionless parameter of the model 
characterizing the total optical depth of the disc.

Dividing the hydrostatic equilibrium equation (equation~(\ref{eq.hydrostat}) or the first equation in (\ref{eq.Pis})) through 
the opacity coefficient and using the relation $\di\tau = -\varkappa \, \rho \, \di z$, we obtain:
\begin{equation}
    \frac1{\varsigma+1} \frac{\partial P^{\varsigma+1}}{\partial \tau} = \frac{\Omega^2 \, z_0 \, \R \, T^{\psi+\varsigma}}{\varkappa_0 \, \mu} \,.
\end{equation}
Near the photosphere the coordinate $z(\tau)\approx z_0$ is almost constant.
Integrating the last equation from $\tau = 0$ to $\tau = 2/3$ yields the boundary condition for dimensionless pressure:
\begin{equation}
 \begin{split}
    p|_{\sigma = 1} =& \left\{
        \frac{3(\varsigma+1)}{16 \cdot 2^{(\psi+\varsigma)/4}}
        \frac{\Pi_1 \Pi_2}{\Pi_4}
        \left( \frac{16}{3} \frac{\Pi_4}{\tau_0} \right)^{(\psi+\varsigma+4)/4}
        f\left(\frac23\right)
        \right\}^{1/(\varsigma+1)}\,,\\
    f(\tau) \equiv& \int_0^{\tau}{\left( 1 + \frac32 \tau' \right)^{(\psi+\varsigma)/4} \, \di\tau'} \,.
 \end{split}
\end{equation}

\begin{figure*}
 \centering
 \includegraphics[width=0.84\textwidth]{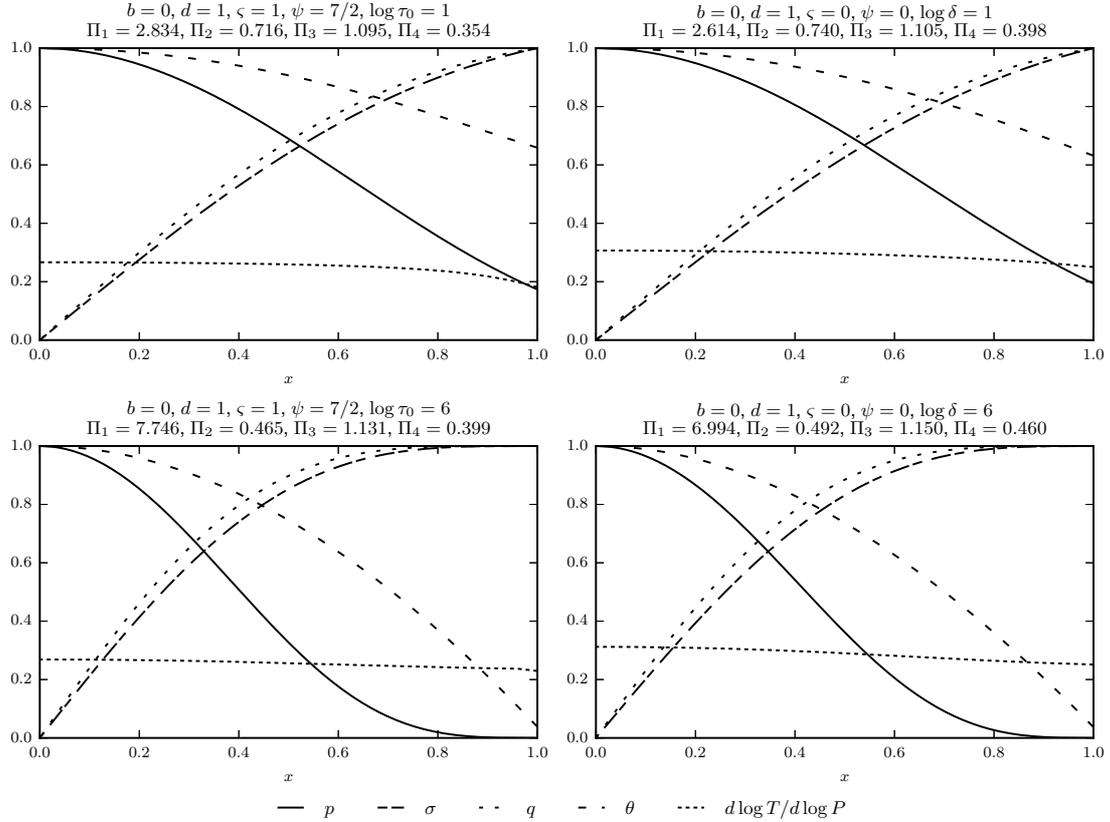}
 \caption{Vertical distribution of various dimensionless variables in a Keplerian $\alpha$-disc. Four dimensionless functions $p(x), \sigma(x), q(x), \theta(x)$ and the derivative $\di\log T / \di\log P$ are shown. The value of $\di\log T / \di\log P$ for all cases considered is below $1/c_P=2/5$ so the convective instability condition~\eqref{eq.conv_cond} is never met. 
Left-hand plots show the case of Kramer's opacity (the 'region c' in the standard $\alpha$-disc theory) 
with the free parameter $\tau_0 = 10^1$ (upper panel) and $\tau_0 = 10^6$ (bottom panel). Right-hand plots show 
the case of Thomson scattering opacity (the 'region b' in the standard $\alpha$-disc theory) with the free parameter $\delta = 10^1$ (upper panel) and $\delta = 10^6$ (lower panel). }
\label{fig.alpha}
\end{figure*}

\begin{figure*}
 \centering
 \includegraphics[width=0.84\textwidth]{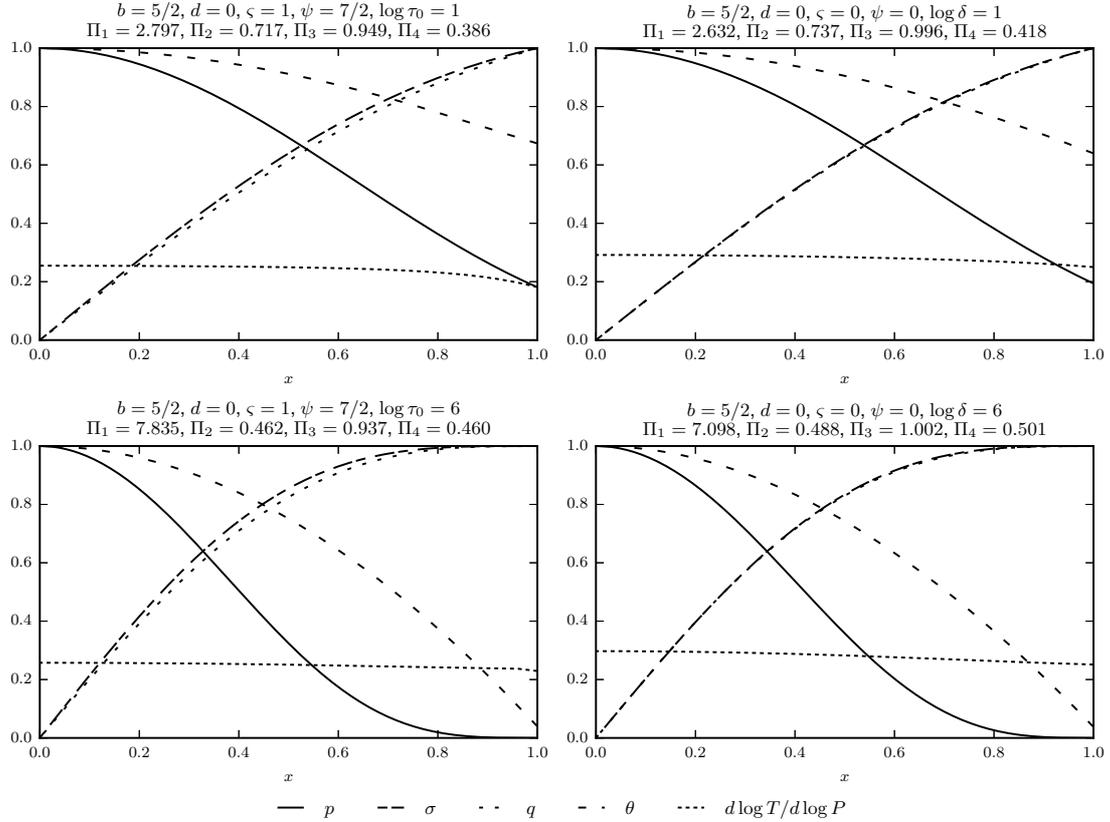}
 \caption{The same as in Fig.~\ref{fig.alpha} for a Keplerian 
disc with ion viscosity $\eta~=~\eta_\centr~\theta^{5/2}$.  
The two right-hand panels with Thomson scattering dominated opacity show that the energy flux is $q~\approx~\sigma$ and $\Pi_3 \approx 1$, suggesting a polytropic-like structure with polytrope index $n \approx 5/2$: $\lambda \approx \theta^{5/2}$.}
\label{fig.ion}
\end{figure*}

\subsubsection{Scattering dominated atmosphere}
\label{s:scattering}

If the opacity is dominated by Thomson scattering, the thermalization of photons occurs at 
the effective optical depth $\taueff$, which is defined as 
\begin{equation}
    \di \taueff = - \sqrt{\varkappa_\mathrm{ff} \, \varkappa_{T}} \, \rho \, \di z \,,
\end{equation}
where $\varkappa_\mathrm{ff} \ll \varkappa_\mathrm{T}$ is the true absorption coefficient, $\varkappa_\mathrm{T} \simeq \varkappa = \varkappa_\centr = \varkappa_0$ is the Thomson scattering opacity.
Thermalization of radiation takes place at the effective optical depth about one, and therefore
\begin{equation}
    T|_{\taueff \approx 1} = T_\eff \,.
\end{equation}
The actual optical depth $\tau$ is determined by scattering: $\di \tau = -\varkappa \, \rho \, \di z$.
In the scattering dominated case, at the photosphere $\tau_{\taueff = 1} \gg 1$.
Therefore, the boundary condition at the photosphere for the dimensionless temperature $\theta$ is
\begin{equation}
    \theta_{\sigma = 1} =
        \left( \frac{8}{3} \frac{\Pi_4}{\varkappa \, \Sigma} \left( 1 + \frac32 \tau \right) \right)^{1/4} \simeq
        \left( \frac{4 \, \Pi_4 \, \tau_{\taueff = 1}}{\varkappa \, \Sigma} \right)^{1/4} \,.
\end{equation}
Correspondingly, the boundary condition at the photosphere for the dimensionless pressure is
\begin{equation}   p_{\sigma = 1} = \Pi_1 \Pi_2 \frac{\tau_{\taueff=1}}{\varkappa \, \Sigma} \,.
\end{equation}

Finally, we find it convenient to introduce the new free parameter $\delta$ 
as the ratio of half the total optical depth to the optical depth at the thermalization depth:
\begin{equation}
    \delta \equiv \frac{\varkappa \, \Sigma}{\tau_{\taueff=1}} \,.
\end{equation}
This parameter will be used below to characterize the flow in the scattering dominated case. 

\subsection{Solution of vertical structure equations~\eqref{eq.Pis}}

The system of differential equations~\eqref{eq.Pis} with boundary conditions described in Section~\ref{s:boundcond} was solved numerically.
The corresponding \textsc{python}-code with use of \textsc{scipy} module
\citep{scipy,zhu1997algorithm,dierckx1995curve} can be freely downloaded from the authors' home page\footnote{\url{http://xray.sai.msu.ru/~malanchev/convinstab/}}.

We use the logarithmic form of the convection stability condition~\eqref{eq.schwarzschild} for the perfect monoatomic gas:
\begin{equation}
    \frac{\di \log T}{\di \log P} \geq
        \left( \frac{\di \log T}{\di \log P} \right)_\mathrm{ad} = \frac{1}{c_P} = \frac25 \,.
\label{eq.conv_cond}
\end{equation}
We found that $\di \log T / \di \log P$ reaches maximum in the disc symmetry plane $x=\sigma=0$.
This value can be found from~\eqref{eq.Pis}:
\begin{equation}
    \left( \frac{\di \log T}{\di \log P} \right)_\centr = \frac{\Pi_3 \Pi_4}{\Pi_1 \Pi_2^2} \,.
\end{equation}

To compare our results with \citet{1998A&AT...15..193K},  we computed the vertical structure of an $\alpha$-disc ($b=0$, $d=1$ 
in the dynamical viscosity prescription \eqref{eq.eta_tp}).
We found that both absorption dominated and scattering dominated $\alpha$-discs are convectively stable for any values of free parameters~$\tau_0$ and~$\delta$.
Figure~\ref{fig.alpha} shows the vertical distribution of dimensionless functions $p, \sigma, q, \theta$ for $\alpha$-disc with Kramer's and Thomson opacities for two values of $\tau_0$ and $\delta$. 

Next we examine an optically thick Keplerian disc with ion viscosity $b=5/2$, $d=0$ (see Fig.~\ref{fig.ion}).
In the Thomson scattering dominated case (right-hand panels of Fig.~\ref{fig.ion}) the disc has a nearly polytropic 
structure with polytrope index $n\approx 5/2$.
The disc is found to be convectively stable.
Clearly, for sufficiently strong dependence of the viscosity (heat generation) on temperature characterized by high power index $b>b_{\mathrm{crit}}$ in \eqref{eq.eta_tp}, the flow should become convectively unstable.
%
We find that for $d = 0$ and Thomson scattering, the critical value $b_\mathrm{crit} \approx 7.97$, for Kramer's opacity $b_\mathrm{crit} \approx 18.54$. 

Note that the opacity coefficient in general form~\eqref{eq.varkappa} has been used in many studies of convection disc stability. The most known is the instability condition 
$2\psi < 3 (\varsigma - 1)$ (see \citet{1980MNRAS.191..135T,1980MNRAS.191...37L}), which is met 
if hydrogen is partially ionized \citep{1994ApJ...427..987B}. 
For direct dependence of the opacity on temperature (i.e. negative $\psi$) the thermal disc instability takes place \citep{hoshi1979,meyer_meyer-hofmeister1981,smak1982}. However, in this paper we restrict ourselves to considering only cases with ion viscosity and absorption or scattering dominated opacities discussed above.

\section{Discussion}
\label{s:discus}

When calculating the vertical structure of laminar shear flows with viscous heating and heat thermal conductivity 
in Section \ref{s:T(z)}, a superadiabatic temperature gradient 
was found to occur at some height above the disc plane. We have used the local Schwarzschild criterion for convection 
\eqref{eq.schwarzschild} to occur. However, more general criterion for convection requires the dimensionless Rayleigh number 
Ra to be large. The Rayleigh number compares the effect of buoyancy forces and dissipation processes and can be determined as     
\beq{}
\mathrm{Ra}=\frac{g_z \, |\partial T/\partial z| \, z_0^4}{\nu \, {\cal K} \, T}
\eeq
where $g_z=\Omega^2z$ is the vertical gravity acceleration, 
$\nu \equiv \eta/\rho$ is the kinematic viscosity, ${\cal K}=\kappa/(\rho {\cal R} c_P/\mu)$ is the thermal diffusivity. 
Making use of $(dT/dz)/T\sim 1/z_0$ and definition \eqref{eq.xi}, 
the Rayleigh number can be expressed through the Reynolds number $\mathrm{Re}=\vs z_0/\nu$ and the Prandtl number Pr \eqref{eq.Prandtl} as
\beq{}
\mathrm{Ra}=\xi^2 \, \mathrm{Pr} \,\mathrm{Re}^2\,.
\eeq
For the typical values of the problem $\xi>1$ and $\mathrm{Pr}\gtrsim 10^{-2}$, Ra turns out to be very large for typical high Reynolds numbers 
for laminar gaseous flows. Therefore, the appearance of convection due to ion viscous heating is possible. 

In the upper convective layer of the flow discussed in Section \ref{s:Sect2}, we have assumed adiabatic convection \eqref{eq.conv_dt_dx}. In fact, the adiabaticity may not hold near the surface because of low density, i.e. the temperature gradient can be higher than 
the adiabatic one, which would decrease the size of the convective zone and the disc thickness $z_0$. However, the 
convective layer can hardly disappear since the temperature in the transition point is found to be rather high, $\sim 0.4$ of the central temperature (see Fig.~\ref{fig.theta_x}). 

In the second part of the paper, we have considered vertical structure of optically thick Keplerian discs
with ion viscosity heating and radiation energy transfer. It is easy to see that in this case the radiation
energy flux is higher than the electron heat conductivity  flux. Indeed, the energy flux due to 
electron heat conductivity is 
$Q_\mathrm{e}=-\kappa dT/dz$ and the radiative flux is $Q_\gamma=-c/(3\varkappa)d(a_rT^4)/dz$, and their ratio is 
\beq{}
\frac{Q_\gamma}{Q_\mathrm{e}}=\frac{4}{3}\frac{\ar T^3}{\kappa \, \varkappa}\simeq 3 \times 10^4\myfrac{\varkappa}{\varkappa_T}^{-1}\myfrac{T}{1\,\mathrm{eV}}^{1/2}\,,
\eeq
where we have used the heat conductivity coefficient for fully ionized gas from \cite{1962pfig.book.....S}.
Clearly, in optically thick fully ionized discs this ratio is much larger than one, and electron heat conductivity can be 
neglected.

\section{Summary and conclusions}

In this paper we have calculated the vertical structure of steady-state thin Keplerian accretion discs. 
The microscopic ion viscosity is assumed to be the only source of heat generation.
We considered two cases of vertical energy transfer --- due to electron heat conductivity 
in the optically thin discs and due to radiation conductivity in the optically thick discs. 

In the optically thin case, if the microscopic transport coefficients are functions 
of temperature only, the vertical temperature distribution can be calculated from the energy balance equation.
Assuming power-law dependence of these coefficients on temperature~\eqref{eq.kappa}, \eqref{eq.eta}, we solved the energy balance equation~\eqref{eq.difeq} to obtain the vertical temperature gradient~\eqref{eq.u}.
If the surface temperature is small enough, the temperature gradient~\eqref{eq.u} exceeds the adiabatic value at some point, suggesting the appearance of an upper convective layer.
If the Prandtl number exceeds some critical value, $\Pr \geq \Pr_\mathrm{crit}$, the entire disc become convectively unstable.
For Keplerian discs $\Pr_\mathrm{crit}=4/9$.
Solution~\eqref{eq.x_t} also enabled us to calculate the vertical density distribution from the hydrostatic equation, which turned out to be non-polytropic in general case. However, in the special case of constant transport coefficients in a fully laminar disc (at $\Pr<\Pr_\mathrm{crit}$) the vertical density distribution is polytropic.      

For optically thick stationary Keplerian thin discs with radiative energy transfer, the vertical structure is calculated from system of equations~\eqref{eq.Pis} supplemented with eight boundary conditions (see Section~\ref{s:boundcond}).
Two boundary conditions for temperature are set at the photosphere, location of which is found for two opacity laws --- absorption dominated atmosphere (Section~\ref{s:absorption}) and Thomson scattering dominated atmosphere (Section~\ref{s:scattering}).
For completeness, we calculate the vertical structure of standard $\alpha$-discs, which was earlier 
considered by \citet{1998A&AT...15..193K}. These discs are found to be convectively stable (see Fig.~\ref{fig.alpha}). 
Their vertical structure generally cannot be described by a polytrope.  
Optically thick Keplerian discs with ion viscosity and electron heat conductivity are found to be convectively stable
for both opacity laws. A polytropic-like structure with polytrope index $n\approx 5/2$ is recovered for discs with Thomson scattering dominated atmospheres (see Fig.~\ref{fig.ion}, right-hand panels).
The four dimensionless parameters $\Pi_{1..4}$ of the vertical disc structure 
determined from the solution of equations~\eqref{eq.Pis}
are needed to calculate the radial disc structure \citep{2007ARep...51..549S}.

The appearance of convection in laminar Keplerian discs can cause turbulence (see, e.g., \cite{2010MNRAS.404L..64L}), which is required for efficient angular momentum transfer. 
In the convectively stable cases, the vertical structure of laminar flows calculated in this paper can be used
as a background solution for further analysis of evolution of small hydrodynamic perturbations, which 
will be considered elsewhere.

\section*{Acknowledgements}
We thank the anonymous referee for critical remarks.
This work is supported by the Russian Science Foundation grant 14-12-00146.

\bibliographystyle{mnras}
\expandafter\ifx\csname natexlab\endcsname\relax\def\natexlab#1{#1}\fi
\bibliography{convinstab}


\end{document}